\documentclass[traditabstract,printer]{aa} 
\usepackage{graphicx,natbib}
\usepackage{txfonts}
\begin{document}
   \title{Upper limit for the D$_2$H$^+$ ortho-to-para ratio in the prestellar core 16293E (CHESS)\thanks{The chemical network is available in electronic form at the 
   CDS via anonymous ftp to cdsarc.u-strasbg.fr (130.79.128.5) or via http://cdsweb.u-strasbg.fr/cgi-bin/qcat?J/A+A}}

   \author{C. Vastel\inst{1,2}, P. Caselli\inst{3}, C. Ceccarelli\inst{4}, A. Bacmann\inst{4}, D.C. Lis\inst{5}, E. Caux\inst{1,2}, 
   C. Codella\inst{6},
   J. A. Beckwith\inst{7}, T. Ridley\inst{3}          }

   \institute{Universit\'e de Toulouse, UPS-OMP, IRAP, Toulouse France \\
         \and
             CNRS, Institut pour la Recherche en Astrophysique et Plan\'etologie, 9 Av. Colonel Roche, BP 44346, 31028 Toulouse Cedex 4, France\\
           \email{charlotte.vastel@irap.omp.eu}
           \and
           	School of Physics and Astronomy, University of Leeds, Leeds LS2 9JT, UK
	  \and
	  	Université Joseph Fourier and CNRS, Institut de Planétologie et d'Astrophysique, Grenoble, France 
	   \and
	       California Institute of Technology, Cahill Center for Astronomy and Astrophysics, Pasadena, CA 91125, USA
	   \and
                   INAF Osservatorio Astrofisico di Arcetri, Largo E. Fermi 5, I-50125, Firenze, Italy             
             \and
           	School of Earth and Environment, University of Leeds, Leeds LS2 9JT, UK
             }

   \date{Received ; accepted }

 
  \abstract
   {The H$_3^+$ ion plays a key role in the chemistry of dense interstellar gas clouds where stars and planets are forming. The low temperatures and 
   high extinctions of such clouds make direct observations of H$_3^+$ impossible, but lead to large abundances of H$_2$D$^+$ and D$_2$H$^+$, 
   which are very useful probes of the early stages of star and planet formation. The ground-state rotational ortho--D$_2$H$^+$ 1$_{1,1}$--0$_{0,0}$ transition at 
   1476.6 GHz in the prestellar core 16293E has been searched for with the Herschel\thanks{{\it Herschel} is an ESA space observatory 
   with science instruments provided by European-led Principal Investigator consortia and with important participation from NASA.} HIFI instrument, 
   within the CHESS (Chemical HErschel Surveys of Star forming regions) Key Program. The line has not been detected at the 
   21 mK~km~s$^{-1}$ level (3$\sigma$ integrated line intensity). We used the ortho--H$_2$D$^+$ 1$_{1,0}$--1$_{1,1}$ transition and 
   para--D$_2$H$^+$ 1$_{1,0}$--1$_{0,1}$ transition detected 
   in this source to determine an upper limit on the ortho-to-para D$_2$H$^+$ ratio as well as the para--D$_2$H$^+$/ortho--H$_2$D$^+$ ratio 
   from a non-LTE analysis.  The comparison between our chemical modeling and the observations suggests that the CO depletion must 
   be high (larger than 100), with a density between 5 $\times$ 10$^5$ and 10$^6$ cm$^{-3}$. Also the upper limit on 
   the ortho--D$_2$H$^+$ line is consistent with a low gas temperature ($\sim$ 11 K) with a ortho-to-para ratio of 6 to 9, i.e. 2 to 3 times higher than the value 
   estimated from the chemical modeling, making it impossible to detect this high frequency transition with the present state of the art receivers.}
   

   \maketitle
%

\section{Introduction}

In the recent years, the chemistry of dark clouds and star forming regions has constantly been revised with the discovery of multiply deuterated
molecules: D$_2$CO \citep{Turner1990,Ceccarelli1998}, ND$_2$H \citep{Roueff2000}, D$_2$S \citep{Vastel2003}, 
ND$_3$ \citep{Lis2002,vandertak2002}, 
CHD$_2$OH, CD$_3$OH \citep{Parise2002,Parise2004} and D$_2$H$^+$ \citep{Vastel2004}. 
Two main pathways can be invoked for understanding the observed large
deuterium fractionation. The first is based on grain chemistry \citep{Tielens1983}. \citet{Vastel2003} and \citet{Parise2004} 
showed that grain chemistry models require a very high atomic D/H ratio accreting on the grains in order to explain their high 
deuterium fractionation ratio that could not be reproduced by gas-phase modeling at that time. New models can now reproduce the 
observed abundance of the multi-deuterated isotopologues of formaldehyde and methanol, both formed in the last stage of the prestellar phase 
\citep{Taquet2012}. 
The second pathway for forming deuterated molecules is based on gas-phase chemistry and results from the ion-molecule
deuterium exchange reactions taking place at low temperatures. In this scenario, deuterated molecules are produced through 
successive reactions starting with H$_2$D$^+$, dominant at temperatures lower than 20 K, CH$_2$D$^+$ or C$_2$HD$^+$, 
dominant at higher temperatures \citep{Roberts2000,Gerlich2002}. H$_2$D$^+$ has proven to be a very good probe 
of the dense cold cores, where CO disappears from the gas phase and is depleted onto the dust grains \citep{Caselli2003,Caselli2008}. 
\citet{Phillips2003} pointed out that the deuteration should be extended beyond H$_2$D$^+$, to D$_2$H$^+$ and D$_3^+$
and suggested that the detection of the D$_2$H$^+$ ion might be possible.
Calculations including all possible deuterated isotopomers of H$_3^+$ have confirmed that, in dense CO depleted regions, the abundance 
of D$_2$H$^+$ will be similar to that of H$_2$D$^+$,
and the D/H ratio will be largely enhanced \citep{Roberts2003,Roberts2004,Walmsley2004,Ceccarelli2005}.\\
D$_2$H$^+$ in its para form was detected for the first time in the prestellar core 16293E by \citet{Vastel2004}, showing its importance, 
as well as that of D$_3^+$ , in determining the total deuterium abundance in the gas phase. This dense core, revealed by amonia emission 
\citep{Mizuno1990}, is sheltered in the 
dense cloud L1689N \citep[see][]{Wootten1987}, in the Ophiuchi region \citep[distance 120 pc:][]{Loinard2008} and has been revealed by millimeter lines and 
continuum emission. This cloud, also harboring a young binary protostellar object (IRAS 16293-2422 A and B) and bipolar outflows 
(named Rho Oph East by \citet{Fukui1986}), 
has been extensively observed, revealing an extreme molecular deuteration in particular towards the cold dense core, named 16293E by 
\citet{Loinard2001} and \citet{Castets2001}. Note that this core has been called $\rho$ Oph E (as in Rho Oph East referring to the outflow emanating 
from IRAS 16293-2422) by \citet{Saito2000} and \citet{Gerin2001}. However, we decided to use the name 16293E since $\rho$ Oph E already refers 
to another condensation within the L1688 cloud in the Ophiuchi complex \citep{Loren1990} and, therefore, is confusing.
We present here recent Herschel/ HIFI observations obtained within the Herschel guaranteed time Key Program CHESS, and modeling of both 
ground state transitions of 
para-- and ortho--D$_2$H$+$, as well as ortho--H$_2$D$^+$, using the recent collisional coefficients with ortho-- and para--H$_2$ \citep{Hugo2009}.

\begin{figure}[!ht]
\includegraphics[scale=0.35,angle=270]{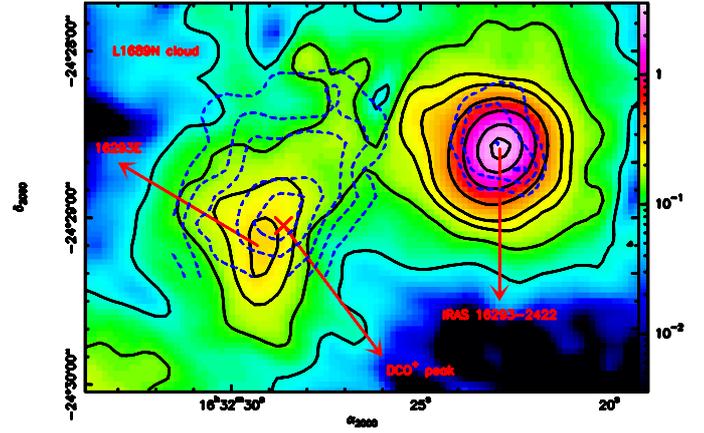}
\caption{Continuum map at 1.3 mm convolved to 15$^"$ with black contours at 4\%, 6\%, 8\%,
15\%, 30\%, 60\%, and 90\% of the peak located at the IRAS 16293-2422 position (6.7 Jy in a 1500 beam). Dashed blue lines show the 
distribution of the DCO$^+$~3--2 integrated 
intensity between 2.8 and 5 km~s$^{-1}$, with contour levels  contour levels are 50\%, 60\%, 70\%, 80\% and 
90\% of the peak (integrated over velocities between 2.8 and 5 km~s$^{-1}$). The cross represent the position towards which 
H$_2$D$^+$ and D$_2$H$^+$ have been observed.}
\label{fig:continuum}
\end{figure}

Figure \ref{fig:continuum} shows the 1.3 mm continuum map as well as the DCO$^+$ contour map (blue) \citep{Lis2002}. 
Studies of the gas kinematics using CO, HCN, H$^{13}$CO$^+$, HCO$^+$ and DCO$^+$ tracers lead to the conclusion that the 
deuterium peak is a part of the ambient cloud that is pushed and compressed by the outflow \citep{Lis2002}. 
Indeed this shock could have released deuterated 
species, that were condensed on the dust grains, into the gas phase with a subsequent cooling of the gas to lower temperatures. This effect, 
combined with the low temperature gas-phase chemistry in the high-density shock-compressed gas leads to 
a high molecular deuteration as observed in: 
DNC \citep[9\%: ][]{Hirota2001}, D$_2$CO \citep[40\% $\pm$ 20\%][]{Loinard2001, Ceccarelli2002}, N$_2$D$^+$ \citep{Gerin2001}, 
NH$_2$D \citep[19\%][]{Mizuno1990,Loinard2001,Roueff2005}, ND$_2$H \citep[4\%: ][]{Mizuno1990,Loinard2001,Lis2006}, DCO$^+$ and 
DCN \citep[$\sim$ 10\%: ][]{Lis2002}, 
HDS \citep{Vastel2003}, D$_2$H$^+$ \citep{Vastel2004}, HDO \citep{Stark2004}, ND$_3$ \citep[0.1\%: ][]{Roueff2005}.
Because no far-infrared  or submillimeter point source has been found, the source can be classified as a prestellar core, the early phase of the 
formation of a protostellar object, before gravitational contraction occurs.

\section{Observations and data reduction}

The para--D$_2$H$^+$ ground state transition was observed towards the prestellar core 16293E with the HIFI instrument \citep{deGraauw2010} 
on board the Herschel Space Observatory \citep{Pilbratt2010}, as part of the Herschel guaranteed time Key Program CHESS \citep{Ceccarelli2010}. 

\begin{figure}[!ht]
\includegraphics[scale=0.5,angle=0]{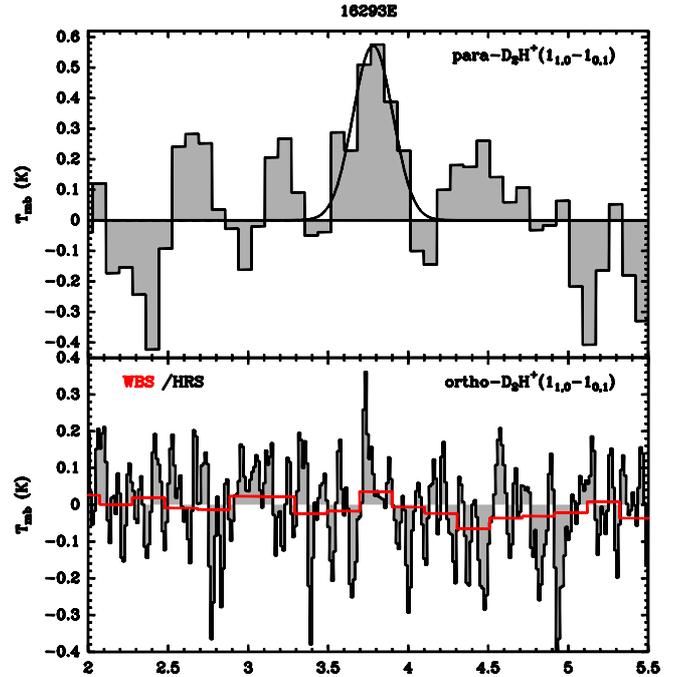}
\caption{para--D$_2$H$^+$ (1$_{1,0}$--1$_{0,1}$) and ortho--D$_2$H$^+$ (1$_{1,0}$--1$_{0,1}$) transitions observed with the CSO and Herschel/HIFI. 
The Wide Band Spectrometer is represented in red, and the High Resolution Spectrometer in black. }
\label{fig:obs}
\end{figure}

A pointing at a frequency centered on the D$_2$H$^+$ line ($\sim$ 1476.6 GHz) with the band 6a HEB receiver was performed on February 16, 
2011 using the pointed Double Beam Switch (DBS) mode with optimization of the continuum. In this mode, both the HIFI Wide Band Spectrometer (WBS) --- providing a spectral resolution of 1.1 MHz ($\sim$ 0.2 km~s$^{-1}$) over an instantaneous bandwidth of 2.4 GHz --- and the HIFI High Resolution Spectrometer (HRS) --- providing a spectral resolution of  125 kHz over an instantaneous bandwidth of 0.12 GHz --- were used. The DBS reference positions were situated approximately 3$^{\prime}$ East and West of the source. The HIFI beam size at the observed frequency is about 14$^{\prime\prime}$, and the main beam and forward efficiencies are about 0.72 and 0.96, respectively \citep{Roelfsema2012}. The data have been processed using the standard HIFI pipeline up to level 2 with the ESA-supported package HIPE 8.0 \citep{Ott2010}. 
The on-source integration time for this observation was 18157 seconds. FITS files from level 2 data were created and translated into CLASS/GILDAS format for subsequent data reduction and analysis. For each scan, a low order polynomial baseline was fitted outside the line window. The antenna temperatures were finally converted to the $T_{mb}$ scale, using the theoretical values of the main beam and forward efficiencies given above.
We present in Figure \ref{fig:obs} this observation as well as the para ground state transition observed at the Caltech Submillimeter Observatory 
(CSO) \citep{Vastel2004}. Both observations were pointed towards the DCO$^+$ peak emission, at coordinates $\alpha_{2000}$ = 16$^h$ 32$^m$ 28$\fs$62,
$\delta_{2000}$ = $-$ 24$\degr$ 29$\arcmin$ 2.7$\arcsec$ \citep[see for example][Figure 9]{Roueff2005}. Note that the coordinates in 
\citet{Vastel2004} were not correctly quoted in the text and should be replaced by the here quoted coordinates.\\

The two nuclear-spin species of H$_2$D$^+$ and D$_2$H$^+$ (ortho and para) are considered separately. 
The parameters of the ortho-- and para--D$_2$H$^+$ and ortho--H$_2$D$^+$ ground transitions are obtained using the 
CASSIS\footnote{CASSIS (http://cassis.irap.omp.eu/) 
has been developed by IRAP-UPS/CNRS.} software, which takes into account the ortho
and para forms separately, with an independent computation of the partition function (cf. Formalism for the CASSIS software, 
http://cassis.irap.omp.eu/), and are reported in Table 1. The Einstein coefficient for the ortho--D$_2$H$^+$ transition ($\sim$ 3 10$^{-3}$ s$^{-1}$) 
is much larger than for the para--D$_2$H$^+$ ($\sim$ 4.6 10$^{-4}$ s$^{-1}$) and ortho--H$_2$D$^+$ ($\sim$ 1.1 10$^{-4}$ s$^{-1}$) transition, 
since it varies as a function of $\nu^3$.
Note that we adopt here the most recent measurement of \citet{Amano2005} of the para-D$_2$H$^+$ line frequency (691.660483 GHz), 372.421385 GHz 
for the ortho--H$_2$D$^+$ transition and 1476.605708 GHz for the ortho--D$_2$H$^+$ transition \citep{Asvany2008}. From a simple Gaussian fitting function, the 
resulting $V_{lsr}$ are (3.78 $\pm$ 0.02) km~s$^{-1}$ for the para--D$_2$H$^+$ transition and (3.59 $\pm$ 0.02) km~s$^{-1}$ for the ortho--H$_2$D$^+$ 
transition (statistical uncertainty from the fit). The minimum difference between the lines is 0.15 km~s$^{-1}$, compatible 
with the $\sim$ 0.1 km~s$^{-1}$ resolution of the acousto-optic spectrometer at 692 GHz, and 0.04 km~s$^{-1}$ at 372 GHz. 
A difference in the observed $V_{lsr}$ is found \citep[exploitation of the data mentioned in][but not published]{Lis2002} in the 
DCO$^+$ 5--4 (3.65 $\pm$ 0.01 km~s$^{-1}$) and 3--2 (3.55 $\pm$ 0.01 km~s$^{-1}$) transitions. The quoted uncertainties correspond to the Gaussian fit 
and the line parameters are listed in Table 1. 
A larger difference in the line center velocity is noticeable comparing the ortho--ND$_2$H 1$_{1,1}$--0$_{0,0}$ (V$_{lsr}$ = 3.64 km~s$^{-1}$) and 
1$_{1,1}$--0$_{0,0}$ (V$_{lsr}$ = 3.60 km~s$^{-1}$) transitions \citep{Lis2006}, as well as the ortho--NH$_2$D 1$_{1,1}$--1$_{0,1}$ 
(V$_{lsr}$ = 3.65 km~s$^{-1}$) and para--NH$_2$D 1$_{1,1}$--1$_{0,1}$ (V$_{lsr}$ = 3.67 km~s$^{-1}$) transitions \citep{Loinard2001}, 
with the ND$_3$ 1$_0$--0$_0$ (V$_{lsr}$ = 3.35 km~s$^{-1}$) transition \citep{Roueff2005}.
Such a difference could be explained by dynamical motions in this source that cannot be reproduced with a simple Gaussian fitting function. The interaction 
between the outflow of IRAS 16293-2422 and the cold core 16293E is likely to produce a velocity shift compared to the L1689N parental cloud, with a 
difference between the non deuterated species and their increasingly deuterated forms. The difference in the beam sizes of the numerous observations may 
be responsible for such a variation as well. The single-dish observations do not allow to conclude.

\begin{figure}[!ht]
\includegraphics[scale=0.55,angle=0]{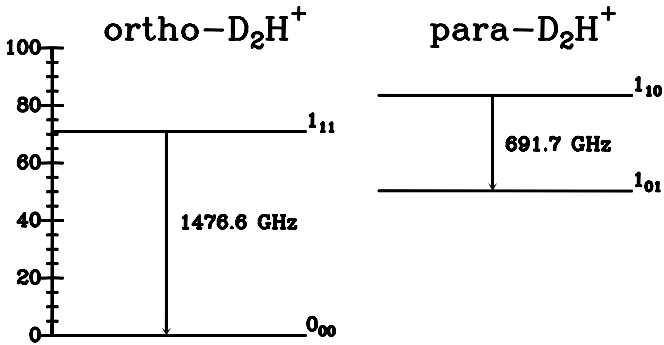}
\caption{Diagram of the lowest energy levels (in Kelvins) of the D$_2$H$^+$ molecule. }
\label{fig:level}
\end{figure}

The 3$\sigma$ upper limit on the integrated line intensity, using the Wide Band Spectrometer, has been derived following the relation:
\begin{center}
\begin{equation}
 3 \sigma~ (\textrm{K~km~s$^{-1}$}) = 3 \times rms \times \sqrt{ 2 \times dv \times FWHM}
\end{equation}
\end{center}
with $rms$ (root mean square) in K, $dv$, the channel width, in km\,s$^{-1}$ and $FWHM$ (Full Width at Half Maximum) in km\,s$^{-1}$. 
We assumed $FWHM = 0.29$\,km\,s$^{-1}$, which is the para--D$_2$H$^+$ emission line width.

\begin{table*}[tb]
  \caption{Derived parameters of the ortho--H$_2$D$^+$ transitions as well as the ortho and para D$_2$H$^+$ fundamental
    lines. DCO$^+$ 3--2 and 5--4 transitions used in \citet{Lis2002} are also quoted using the estimated frequencies from \citet{caselli2005}. 
    Note that a discussion on the V$_{LSR}$ can be found at the end of Section 2.}    
\label{tab:table1}      
\centering                          
\begin{tabular}{c c c c c c c c c c}        
\hline\hline                 
Species      &  Transition & Frequency & Telescope   & beamsize                & $\int T_{mb}$ dV &  rms    &   binsize  & $\Delta$V      & V$_{LSR}$   \\    
	           &                     & GHz            &                      &   ($^{\prime\prime}$) &(mK~km~s$^{-1}$)  & mK     &     (km~s$^{-1}$)   & (km~s$^{-1}$)            & (km~s$^{-1}$)                        \\
\hline                        
ortho--H$_2$D$^+$  &1$_{1,0}$--1$_{1,1}$ &   372.421385  & CSO  &   20  & 720            &    80   & 0.080   &  0.36 $\pm$ 0.04   & 3.59 $\pm$ 0.02  \\
para--D$_2$H$^+$  & 1$_{1,0}$--1$_{0,1}$ &   691.660483  & CSO  &  11   &  183            & 170   & 0.080   &  0.29 $\pm$ 0.05   & 3.78 $\pm$ 0.02   \\
ortho--D$_2$H$^+$  & 1$_{1,1}$--0$_{0,0}$ &   1476.605708  & HIFI/HRS  & 14   &    $\le$  30      &   120    & 0.012   & & \\
ortho--D$_2$H$^+$  & 1$_{1,1}$--0$_{0,0}$ &   1476.605708  & HIFI/WBS  & 14   &    $\le$  21      &   21    & 0.203   & & \\
\hline
DCO$^+$ & 3--2 &  216.1125822  & CSO & 35  & 3235 & 94 & 0.134 & 0.67 $\pm$ 0.03 & 3.55 $\pm$ 0.01\\
DCO$^+$ & 5--4 &  360.1697783  & CSO & 20  & 528    & 109 & 0.081 & 0.54 $\pm$ 0.03 & 3.65 $\pm$ 0.01\\
\hline                                   
\end{tabular}
\end{table*}

\section{Discussion}

\subsection{Line widths}
\label{sec:linewidth}

The line widths are listed in Table \ref{tab:table1}. The expected thermal line width 
varies with the kinetic temperature (T) as:

\begin{equation}
\Delta v_{\rm T} (km~s^{-1}) = 2 \sqrt{2ln2} \times \sqrt{\frac{kT}{m}} ,
\end{equation}
where k is the Boltzman constant and m is the molecular weight. The ortho--H$_2$D$+$ line width corresponds to a 11.3 K 
gas, comparable with the 9.2 K temperature found for para-D$_2$H$^+$. 
However, the uncertainty 
in the fit of the line, using the Levenberg-Marquardt method, should be taken into account leading a temperature of 11.3 $\pm$ 1.8 K for 
H$_2$D$+$ and 9.2 $\pm$ 3.2 K for D$_2$H$+$. The resulting range for the kinetic temperature is therefore [9.5 -- 12.4] K taking 
into account both ions.\\
The observed (o-p) ND$_2$H line width \citep[about 0.45 km~s$^{-1}$ for the 1$_{1,1}$--0$_{0,0}$ transitions;][]{Lis2006}, 
ND$_3$ \citep[0.44 km~s$^{-1}$;][]{Roueff2005} as well as the DCO$^+$ 
line width \citep[about 0.5 km~s$^{-1}$ for the 5--4 transition, less optically thick than the 3--2 transition;][]{Lis2002} and the 
N$_2$D$^+$ line width \citep[0.32 km~s$^{-1}$ for the 3--2 transition;][]{Gerin2001} are systematically larger than the 
thermal line widths of 0.16, 0.12 and 0.16 km~s$^{-1}$ respectively for a 10 K kinetic temperature. Considering the high 
critical densities for the (o-p) ND$_2$H and ND$_3$ transitions (larger than 10$^6$ cm$^{-3}$), it seems that systematic 
motions occur even in the dense part of the cloud. Also, at large enough densities/depletions, one expect to have the light ions 
left as the main tracers of the gas. This further confirms that H$_2$D$^+$ 
and D$_2$H$^+$ remain the only tracers of the cold, dense and CO/N$_2$ depleted central region where other molecular tracers are largely 
condensed onto the dust grains, rather than the lower density envelope when turbulence takes over.

\subsection{H$_2$D$^+$ and D$_2$H$^+$ non-LTE modeling}

\begin{figure}[!ht]
\includegraphics[scale=0.5,angle=0]{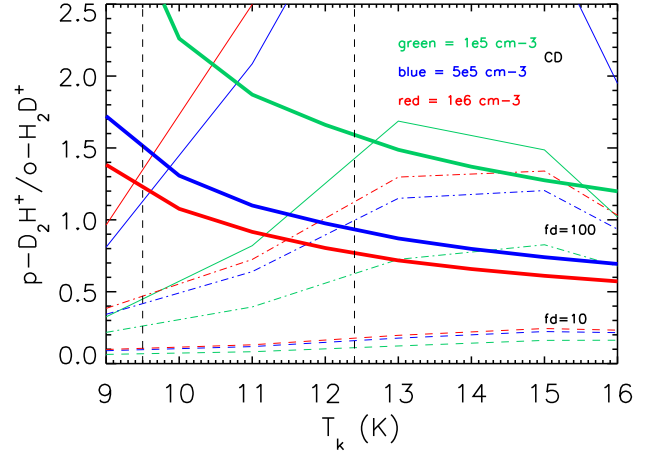}
\caption{Variation of the para D$_2$H$^+$ to ortho H$_2$D$^+$ ratio as a function of the gas temperature, for densities of 
1 $\times$ 10$^5$ (green), 5 $\times$ 10$^5$ (blue) and 1 $\times$ 10$^6$ (red) cm$^{-3}$. The thick lines show the 
ratio derived from observations, using non-LTE modeling, whereas the thin lines show the chemical model results. Three depletion factors at 
steady-state are presented:
complete depletion (CD: solid lines), depletion factor = 100 (dot-dashed lines) and depletion factor = 10 (dashed lines). The thin 
dashed vertical lines present the [9.5 -- 12.4] K range.}
\label{fig:d2hp_h2dp}
\end{figure}

\begin{figure}[!ht]
\includegraphics[scale=0.5,angle=0]{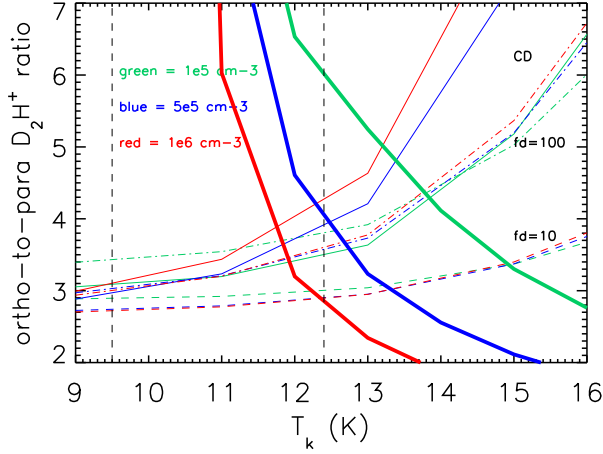}
\caption{Variation of the ortho-to-para D$_2$H$^+$  ratio as a function of the gas temperature, for densities of 
1 $\times$ 10$^5$ (green), 5 $\times$ 10$^5$ (blue) and 1 $\times$ 10$^6$ (red) cm$^{-3}$. The thick lines show the ratio derived from 
observations, using non-LTE modeling, and can be considered as upper limits, whereas the thin lines show the chemical model results. 
Three depletion factors at steady-state are presented:
complete depletion (CD: solid lines), depletion factor = 100 (dot-dashed lines) and depletion factor = 10 (dashed lines). The thin 
dashed vertical lines present the [9.5 -- 12.4] K range. }
\label{fig:d2hp_opr}
\end{figure}

The H$_2$ density is a critical parameter for the interpretation of our deuterated ions observations. CO observations of the 
3--2, 4--3, 6--5 transitions give a lower limit for the molecular hydrogen density of 5 $\times$ 10$^3$ cm$^{-3}$ \citep{Lis2002}. 
In their analysis, \citet{Lis2002} used a kinetic temperature of 12 K \citep[based on NH$_3$ observations of][]{Menten1987} 
and a density of 5 $\times$ 10$^5$ cm$^{-3}$, consistent with their observed DCO$^+$ (5--4)/(3--2) and N$_2$D$^+$ (4--3)/(3--2) 
line ratios. From 450 and 850 $\mu$m continuum maps \citet{Stark2004} inferred an isothermal dust temperature 
T = 16 K and peak density of 1.6 $\times$ 10$^6$ cm$^{-3}$. Note however that their dust peak emission does not correspond 
to the peak of deuterated molecules (about 15" away) that we are studying in the present paper. This could be due to the release 
of the deuterated species that were condensed on the dust grains by the interaction between the outflow from IRAS 16293-2422 and 
the cold core 16293E. This compression is then likely to cool the gas to lower temperatures. These observations 
possibly trace the parental cloud of the prestellar core 16293E. 
We will therefore consider in the following study densities 
between 10$^5$ and 10$^6$ cm$^{-3}$ and kinetic temperatures between 9 and 16 K (see section \ref{sec:linewidth}).\\
From the recent computations of the collision rates \citep{Hugo2009}, the critical densities for the ortho--H$_2$D$^+$, 
ortho--D$_2$H$^+$ and para--D$_2$H$^+$ transitions are $\sim 1.3 \times 10^5$ cm$^{-3}$,  $\sim 7 \times 10^6$ cm$^{-3}$ 
and $\sim 5.6 \times 10^5$ cm$^{-3}$ respectively, using the Einstein coefficients from \citet{Ramanlal2004}, 
1.2 $\times$ 10$^{-4}$, 3.3 $\times$ 10$^{-3}$ and 5.1 $\times$ 10$^{-4}$ s$^{-1}$ respectively. 
From the large difference between the critical densities, the ortho--D$_2$H$^+$ transition will trace denser regions 
than the para transition.
Since the typical density in the core seems to be less than these critical 
densities, a Local Thermodynamic Equilibrium approximation is not applicable. In \citet{Vastel2004}, only LTE 
modeling could be performed because the collision rates were not available at that time.
We produced collisional files, assuming a simple 2 level system, using the inelastic state-to-state rate 
coefficients for the ortho and para ground transitions of H$_2$D$^+$ and D$_2$H$^+$ in collision with para and ortho 
H$_2$, as a function of temperature. 
We used the non-LTE radiative transfer code RADEX \citep{vandertak2007} in the large velocity gradient (LVG) approximation 
with the collisional files that were created for these species.
We could consider in the following non-LTE modeling 
two values for the H$_2$ ortho to para ratio of 3 (highest value at thermodynamic equilibrium) and 0 (only collisions with para 
H$_2$). However, the similarities of the collision coefficients with ortho and para H$_2$ for (o,p)H$_2$D$^+$ or (o,p)D$_2$H$^+$ 
will lead to a very small difference in the derived column densities.  \\
The variation of the para--D$_2$H$^+$ to ortho--H$_2$D$^+$ ratio as a function of the gas temperature, for densities of 
1 $\times$ 10$^5$ (green), 5 $\times$ 10$^5$ (blue) and 1 $\times$ 10$^6$ (red) cm$^{-3}$, are presented in Figure \ref{fig:d2hp_h2dp} 
(thick lines) and the upper limit on the ortho-to-para D$_2$H$^+$ ratio is presented in Figure \ref{fig:d2hp_opr} (thick lines). 
Note that a selection on the optical depth has been performed ($\tau <$ 100), since the excitation temperature found using RADEX 
may then not be representative of the emitting region. The optical depth for the ortho--D$_2$H$^+$ transition is lower than 100 for 
kinetic temperatures larger than $\sim$ 10.5 K.

\begin{figure}[!ht]
\includegraphics[scale=0.5,angle=0]{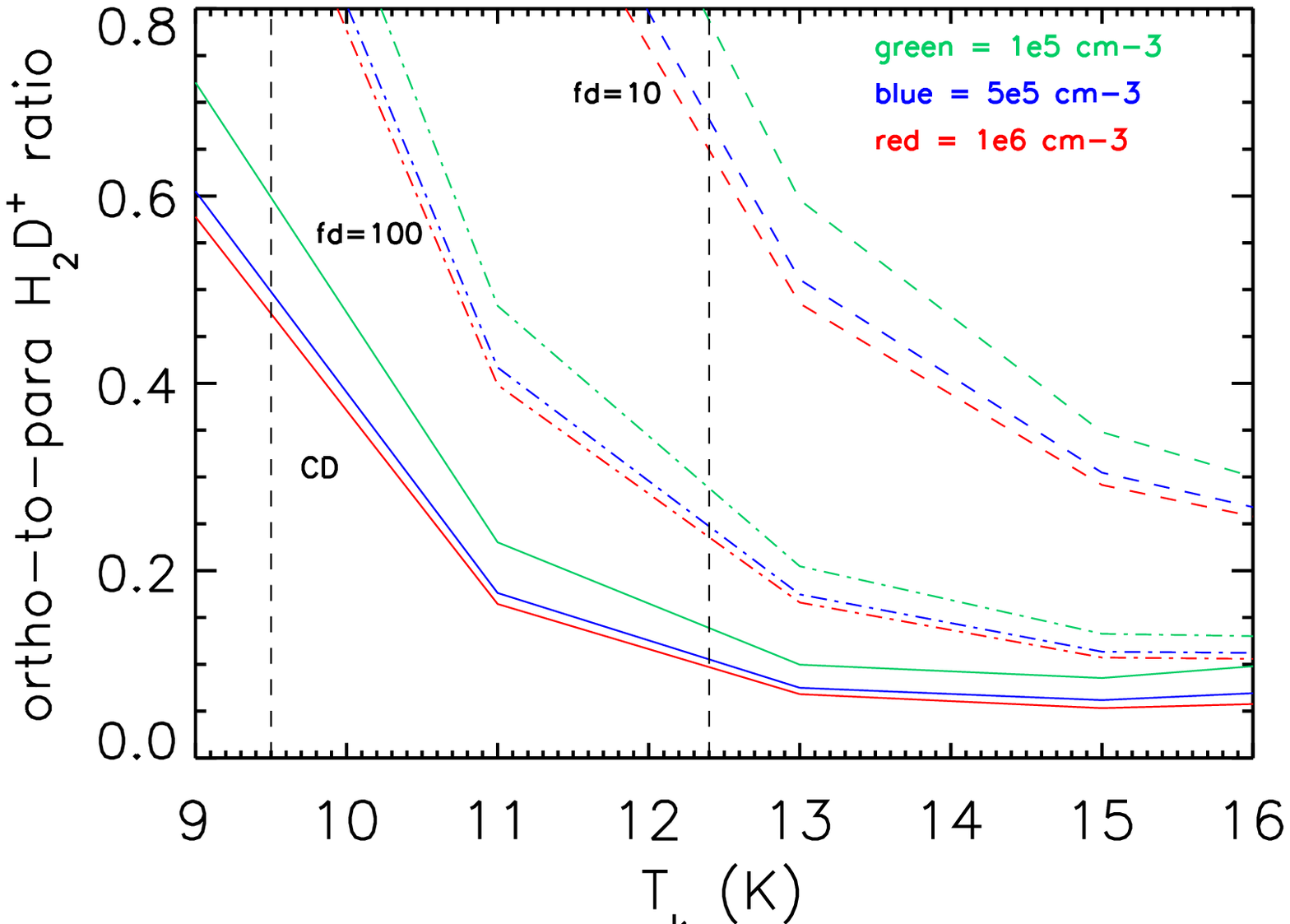}
\caption{Prediction of the ortho-to-para H$_2$D$^+$  ratio as a function of the gas temperature, for densities of 
1 $\times$ 10$^5$ (green), 5 $\times$ 10$^5$ (blue) and 1 $\times$ 10$^6$ (red) cm$^{-3}$. Three depletion factors at steady-state are presented:
complete depletion (CD: solid lines), depletion factor = 100 (dot-dashed lines) and depletion factor = 10 (dashed lines). The thin 
dashed vertical lines present the [9.5 -- 12.4] K range. }
\label{fig:h2dp_opr}
\end{figure}

\subsection{Full chemical modeling}

The chemical network used here has been collected from various sources. First, a reduced chemical network based on the Nahoon code was obtained from the KIDA \footnotemark \footnotetext{http://kida.obs.u-bordeaux1.fr/models} database \citep{Wakelam2012}. The network contains species with three or less elements and includes  hydrogen, helium, carbon, nitrogen and oxygen. The reduced Nahoon code has then been modified to include deuterium and deuterated species, with H$_2$, HD and D$_2$ forming on grain surfaces. Finally, the spin states of all isotopologues of H$_2$ and ${\rm H_3^+}$ were included. The rate coefficients of the new reactions were taken from \citet{Sipila2010} , \citet{Hugo2009}, \citet{Flower2004} and \citet{Walmsley2004}, choosing the most recent values in case of multiple choice.  The reaction balancing routine within the Nahoon program (the one which checks the reaction list, to make sure that reactants and products have the same number of elements and charges) had to be modified after the inclusion of the different spin states,  to avoid to balance spin states between reactants and products (as spin states are not conserved). For this purpose, different spin states were labelled adding an extra column in the element and species definition file. In this column we assigned -1 to the lower spin state and +1 to the higher spin state of all isotopologues of H$_2$ and H$_3^+$ (para-D$_3^+$ has been assigned a value of +2 as it is the highest of three spin states). All other species have a zero in the corresponding column, indicating that no spin state is considered. The H$_2$ self shielding data has not been modified but new parameters for the self shielding of HD and D$_2$ have been introduced and have been each set to initially be equal to that of H$_2$.  Coulomb focusing was taken into account for reactions involving negatively charged ions on neutral grains \citep{Draine1987}.  Only neutral and negatively charged grains are present in the network. The final chemical network includes over 3,500 reactions involving nearly 130 different species. The full chemical network used in this work is available at the CDS via anonymous ftp to 
cdsarc.u-strasbg.fr (130.79.128.5) or via http://cdsweb.u-strasbg.fr/cgi-bin/qcat?J/A+A/.
Comparison with previous work and a parameter space exploration will be presented by Kong et al. (in prep.). 

A range of H$_2$ volume densities (1 $\times$ 10$^5$, 5 $\times$ 10$^5$, 1 $\times$ 10$^6$\,cm$^{-3}$), kinetic temperatures 
(9, 11, 13, 15, and 16 K), and depletion factors ($\equiv$ undepleted abundance / depleted abundance = 10, 100, infinity)  of all the elements 
heavier than helium (C, N and O) have been explored and followed in time until the system reaches equilibrium. 
Since the CO molecule is the main destroyer of the H$_3^+$ ion (and its deuterated counterparts), we simulate three depletion values at steady-state: 
complete depletion, and elemental depletion factor of  10 and 100 corresponding to CO depletion factors of about 14 and 140, 
respectively, at equilibrium. From \citet{Lis2002}, C$^{18}$O observations at the deuterium peak of 16293E lead to a CO depletion factor (average 
along the line of sight) of $\sim$10, representing a lower limit for the central core traced by D$_2$H$^+$.  
Considering the overall parameters in the chemical modeling we adopt the conservative values of 3 $\times$ 10$^{-17}$ s$^{-1}$ for 
the cosmic ionization rate, 0.01 for the dust to gas ratio, and 0.1 $\mu$m for the grain radius. 

\begin{figure}[!ht]
\includegraphics[scale=0.5,angle=0]{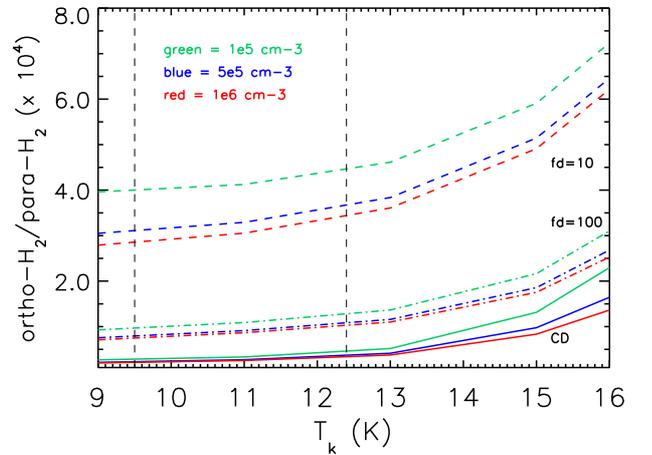}
\caption{Variation of the ortho-to-para H$_2$  ratio as a function of the kinetic temperature, for densities of 
1 $\times$ 10$^5$ (green), 5 $\times$ 10$^5$ (blue) and 1 $\times$ 10$^6$ (red) cm$^{-3}$. 
Three depletion factors at steady-state are presented:
complete depletion (CD: solid lines), depletion factor = 100 (dot-dashed lines) and depletion factor = 10 (dashed lines). The thin 
dashed vertical lines present the [9.5 -- 12.4] K range.}
\label{fig:h2ratio}
\end{figure}

Figure 4--6 show the predictions (thin lines) of the model at 
equilibrium (reached in $2~10^5$ years 
for the complete depletion case, $5~10^5$ years for a depletion factor of 100 and $2~10^6$ years for a depletion factor of 10) for temperatures between 
9 and 16 K for three densities. The para--D$_2$H$^+$/ortho--H$_2$D$^+$ ratio increases with the depletion factor: indeed 
the disappearance of the CO molecule from the gas phase leads to the reactions of HD with H$_3^+$ with an increasing 
production of its deuterated counterparts \citep{Phillips2003}. Also the increase of the ortho-to-para D$_2$H$^+$ ratio with the 
depletion factor can be explained by the fact that the reactions with HD, ortho--D$_2$ and para--D$_2$ converting para--D$_2$H$^+$ into 
ortho--D$_2$H$^+$ are dominant compared to the reverse reactions \citep{Hugo2009}. From equations 4 and 5 we can say that a low 
ortho--H$_2$ value leads to a very high D$_3^+$ abundance for densities larger than 10$^5$ cm$^{-3}$. Indeed the 
D$_3^+$ ion is likely to be the dominant ion in such high-density regions: e.g. Figure 3 from \citet{Flower2006} and Figure 2 from \citet{Sipila2010}. 
The ortho-to-para D$_2$H$^+$ ratio consequently increases as the depletion factor increases, as seen in Figure \ref{fig:d2hp_opr}.  
The ortho-to-para H$_2$ ratio should be taken into account in the overall chemistry, since it is critical for the degree of deuteration of H$_3^+$. 
Increasing the depletion factor leads to a local decrease of the CO abundance. Therefore H$^+$ and H$_3^+$ will mainly react with 
ortho--H$_2$ (at the same temperature the backward reaction is negligible), and not CO. This will lead to a high ortho-to-para H$_2$ conversion, 
consequently a lower ortho-to-para H$_2$ ratio as seen in Figure \ref{fig:h2ratio}.
Our chemical modeling also predicts a variation of the steady-state ortho-to-para H$_2$ ratio for the complete 
depletion case, a depletion factor of 100 and a depletion factor of 10, between 9 and 16 K (see Figure \ref{fig:h2ratio}). 
Because of the large internal energy ($\sim$ 170 K) of the lowest ortho--H$_2$ level (J = 1) compared to the temperature range explored 
in this source, the ortho--H$_2$ form is a limiting factor for deuteration. It overcomes the energy barrier, leading to exothermic (i.e fast and 
temperature independent) reactions \citep[e.g.][]{Gerlich2002}:
\begin{equation}
\rm o-H_2D^+ + o-H_2 \longrightarrow (o,p)-H_3^+ + HD
\end{equation}
\begin{equation}
\rm p-D_2H^+ + o-H_2 \longrightarrow p-H_2D^+ + HD
\end{equation}
The excited nuclear spin state of the D$_3^+$ ion is removed preferentially by ortho--H$_2$ in the following endothermic (by only 18 K) 
reaction:
\begin{equation}
\rm m-D_3^+ + o-H_2 \longrightarrow o-D_2H^+ + HD
\end{equation}
Following \citet{Hugo2009} we have assigned meta--D$_3^+$ with the modification having the lowest ground state energy, corresponding 
to the A1 representation of the symmetry group S3.
All forms should be taken into account in any chemical modeling involving deuterated ions.

Although \citet{Pagani2009} showed that the 
ratio is unlikely in steady-state in prestellar cores, this will not affect the computation of the column densities since, as mentioned in Section 2, 
the collisional rates for ortho--H$_2$D$^+$ (as well as para-- and ortho--H$_2$D$^+$) are similar for collisions with both para-- and 
ortho--H$_2$. Note also that the difference between our modeling and \citet{Pagani2009} comes from the time-scale used. Our 
values are at equilibrium.\\
With a comparison between the chemical model predictions and the observations, three convergences can be found for the 
complete depletion case and $f_d$ = 100. These domains are quoted in Table \ref{tab:abundances} for densities of 10$^5$, 5 $\times$ 10$^5$ 
and 1 $\times$ 10$^6$ cm$^{-3}$. From the gas kinetic temperature range (9.5 -- 12.4 K) found from the uncertainties on the 
para--D$_2$H$^+$ and ortho--H$_2$D$^+$ linewidths, the n$_{H_2}$ = 10$^5$ cm$^{-3}$ domain can be ruled out. 
Should we consider the highest gas kinetic temperature (12.4 K) the modeling results in a para--D$_2$H$^+$/ortho--H$_2$D$^+$ abundance 
ratio 12\% lower than the observations for a molecular density of 10$^5$ cm$^{-3}$. 
This would reduce the upper limit on the ortho-to-para ratio to a value of $\sim$ 6, $\sim$ 40$\%$ higher than the value modeled for 
the complete depletion case and with a depletion factor of 100 (see Figure \ref{fig:d2hp_opr}). Also, the comparison between the chemical modeling and the ortho-to-para 
D$_2$H$^+$ upper limit from the non detection of the ortho 
transition is consistent with a low gas temperature ($<$ 11.7 K for a 10$^6$ cm$^{-3}$ density). A proper model should consider the (unknown) 
physical structure and proper 
time-depended freeze-out.

\begin{table*}[!b]
  \caption{Non-LTE computations for the ortho--H$_2$D$^+$, para-- and ortho--D$_2$H$^+$ column densities for densities 
  of 10$^5$, 5 $\times$ 10$^5$ and 10$^6$ cm$^{-3}$. The temperature range results from the intersection of the modeling 
  (thin lines in Figure \ref{fig:d2hp_h2dp}) and the non-LTE computation (thick lines in Figure \ref{fig:d2hp_h2dp}).}    
\label{tab:abundances}      
\centering                          
\begin{tabular}{c c c c}        
\hline                
n$_{H_2}$ (cm$^{-3}$)	             &  10$^5$  &   5 $\times$ 10$^5$  &  1 $\times$ 10$^6$ \\    
T$_{gas}$	 (K)                                   &  [12.7 -- 15.6]                           & [9.9 -- 12.2]      &  [9.4 -- 11.2]                 \\
\hline                        
~N(ortho--H$_2$D$^+$)       &  [1.4 -- 1.9] $\times$ 10$^{13}$   &  [1.2 -- 1.8] $\times$ 10$^{13}$    &  [1.2 -- 1.9] $\times$ 10$^{13}$  \\
~N(para--D$_2$H$^+$)        &  [1.7 -- 2.9] $\times$ 10$^{13}$    &  [1.1 -- 2.4] $\times$ 10$^{13}$  &  [1.1 -- 2.3] $\times$ 10$^{13}$    \\
~N(ortho--D$_2$H$^+$)       &  $\le$ 1.6 $\times$ 10$^{14}$    & $\le$ 9.2 $\times$ 10$^{14}$   & $\le$ 5.1 $\times$ 10$^{14}$ \\
~[ortho--D$_2$H$^+$]/[para--D$_2$H$^+$]   & $\le$ 6  & $\le$ 44  &   $\le$ 32   \\

\hline                                   
\end{tabular}
\end{table*}

Note that no other molecular tracers like DCO$^+$ and N$_2$D$^+$, observed by \citet{Lis2002} can be used as a comparison with 
the chemical modeling as it might prove difficult to disentangle their contribution from the more extended envelope to 
the central region. \\
Considering an average 11 K gas temperature the upper limit for the ortho-to-para 
D$_2$H$^+$ ratio is 2 to 3 times larger than the value found from the modeling (around 3) for densities larger than 10$^5$ cm$^{-3}$. 
We get, from the non-LTE radiative transfer modeling, $N$(ortho--H$_2$D$^+$) = (1.4 $\pm$ 0.1) $\times$ 10$^{13}$ cm$^{-2}$, 
$N$(para--D$_2$H$^+$) = (1.4 $\pm$ 0.2) $\times$ 10$^{13}$ cm$^{-2}$, and  $N$(ortho--D$_2$H$^+$) $\le$ 1.4 $\times$ 10$^{14}$ cm$^{-2}$. 
These values are reproduced by (or in the case of ortho--D$_2$H$^+$ compatible with) our chemical modeling in the complete depletion case. 
The present state of the art instruments are clearly unable to detect the ortho--D$_2$H$^+$ transition in this source. The CCAT (Cornell Caltech 
Atacama Telescope\footnote{http://www.ccatobservatory.org/}) project appears to be the most adequate to detect this ortho--D$_2$H$^+$ transition 
in warmer regions. \\
From Figure \ref{fig:h2dp_opr}, we can use the average 11 K gas temperature in the complete depletion case as well as with a depletion factor 
of 100 to estimate the radiation temperature of the para--H$_2$D$^+$ transition at 1370.085 GHz for densities larger than 10$^5$ cm$^{-3}$. 
Using a non LTE modeling described in Section 3.2, the resulting para--H$_2$D$^+$ column density is (9.07 $\pm$ 0.49) 10$^{13}$ cm$^{-2}$ 
(T$_R$ = (60 $\pm$ 20) mK) for the complete depletion case, and (3.40 $\pm$ 0.19) 10$^{13}$ cm$^{-2}$  (T$_R$ = (50 $\pm$ 10) mK) for a 
depletion factor of 100. This transition unfortunately falls in a frequency range not covered by the HIFI instrument, but could be a target for future instruments.
Carbon monoxyde in the LDN 1689N cloud is only moderately depleted in the single-dish beam \citep{Lis2002} but it is possible that the angular resolution 
of the existing single-dish data is simply insufficient to show the spatial stratification predicted by our chemical model. 
The presence of completely depleted regions smaller than the CSO beam cannot be ruled out by our observations. \\
The deuterium peak in the LDN 1689N cloud has been argued in the literature to be a shock-compressed interaction region 
between a molecular outflow and an ambient cloud. Therefore high spatial resolution mapping observations of high density tracers, like 
NH$_3$ and its deuterated counterparts as well as N$_2$H$^+$ and N$_2$D$^+$ are necessary to allow investigating the kinematics 
of the high-density gas in this region. Both the Atacama Large Array Millimeter and Expanded Very Large Array are suitable for a follow-up 
of this source that reveals to be a non typical prestellar core. 


 \section{Conclusions}

   \begin{enumerate}
      \item The ground-state ortho--D$_2$H$^+$ 1$_{1,1}$--0$_{0,0}$ transition at 1476.6 GHz in the prestellar core 16293E has been searched for 
      with the Herschel/HIFI instrument. 
      \item The collision rates for the ortho and para H$_2$D$^+$ and D$_2$H$^+$ ions with molecular hydrogen have been used with 
      a non-LTE radiative transfer code (RADEX) to derive the 
      column densities of the detected ortho--H$_2$D$^+$ and para--D$_2$H$^+$ ground transitions as well as the upper limit on the 
      ortho--D$_2$H$^+$ observed with the HIFI instrument on board the Herschel observatory.
      \item We used a gas-phase chemical model, in which deuterium chemistry and the spin states have been added and compared our 
      modeling with the inferred para--D$_2$H$^+$/ortho--H$_2$D$^+$, as well as the upper limit on the ortho-to-para D$_2$H$^+$ 
      ratio. The kinetic temperature is deduced from the line width of both detected ions to be about 10 K, and we varied 
      the molecular hydrogen density between 10$^5$ and 10$^6$ cm$^3$. The upper limit on the observed ortho-to-para D$_2$H$^+$ 
      ratio is consistent with the modeling and points to a low ($\sim$ 11 K) gas kinetic temperature. The detection of the ortho--D$_2$H$^+$ 
      transition in the cold regime is a challenge as an rms of about 
      9 mK is needed, compared to the 21 mK rms reached by the HIFI instrument in about 5 hours.  
   \end{enumerate}

\begin{acknowledgements}
     HIFI has been designed and built by a consortium of
institutes and university departments from across Europe, Canada and the
United States under the leadership of SRON Netherlands Institute for Space
Research, Groningen, The Netherlands and with major contributions from
Germany, France and the US. Consortium members are: Canada: CSA,
U.Waterloo; France: CESR, LAB, LERMA, IRAM; Germany: KOSMA,
MPIfR, MPS; Ireland, NUI Maynooth; Italy: ASI, IFSI-INAF, Osservatorio
Astrofisico di Arcetri-INAF; Netherlands: SRON, TUD; Poland: CAMK, CBK;
Spain: Observatorio Astronomico Nacional (IGN), Centro de Astrobiologia
(CSIC-INTA). Sweden: Chalmers University of Technology - MC2, RSS \&
GARD; Onsala Space Observatory; Swedish National Space Board, Stockholm
University - Stockholm Observatory; Switzerland: ETH Zurich, FHNW; USA:
Caltech, JPL, NHSC. We thank CNES for financial support. We thank Valentine 
Wakelam for providing the reduced Nahoon chemical network and for carefully 
checking the self-consistency of the program. We thank Laurent Loinard for 
kindly providing the data published in 2001. Support for this work was provided by 
NASA  through an award issued by JPL/Caltech. 

\end{acknowledgements}

\bibliographystyle{aa}
\bibliography{biblio_d2hp}

\end{document}